\documentclass[aps,prl,preprint,superscriptaddress]{revtex4}

\usepackage{epsfig} 

\begin{document}

\title{Collective pinning of imperfect vortex lattices  by
material line defects in extreme type-II superconductors}

\author{J. P. Rodriguez}
\thanks{Permanent address: Department of Physics and Astronomy, 
California State University, Los Angeles, California 90032}
\author{M.P. Maley}

\affiliation {Superconductivity Technology Center, 
Los Alamos National  Laboratory, Los Alamos, NM 87545}

\date{\today}

\begin{abstract}
The critical current density shown by a superconductor at the extreme type-II
limit is predicted to  follow a $1/\sqrt B$ power law with external
magnetic field $B$
if the vortex lattice is weakly pinned  by 
material line defects.
It acquires an additional
inverse dependence with thickness along the
line direction once  pinning of the  interstitial vortex lines
by material point defects is included.  
Moderate quantitative
agreement with the critical current density shown by second-generation wires
of high-temperature superconductors 
in kG magnetic fields 
is achieved at liquid-nitrogen temperature.
\end{abstract}

\maketitle

Thin films of the high-temperature superconductor
YBa$_2$Cu$_3$O$_{7-\delta}$ (YBCO)
grown by pulsed laser deposition (PLD) 
on a substrate
that serves to align the crystalline axes
now routinely achieve critical currents that reach a
substantial fraction of the 
maximum depairing current\cite{dam}\cite{klaassen}\cite{suenaga}\cite{maiorov}.
Such films also typically show a peak in
the critical current as a function of the orientation of external
magnetic field 
near the crystallographic $c$ axis,
perpendicular to the film (copper-oxygen) plane\cite{angle}.
It is well known that the Abrikosov vortex lattice that exists
in external magnetic field shows no critical current if it is
not pinned down by material defects\cite{LO}\cite{Tink}.
The $c$-axis peak in the critical current shown by 
films of PLD-YBCO 
therefore indicates that some fraction of the  material defects
that pin the vortex lattice 
must be correlated along the $c$ axis as well.
Careful studies of the 
microstructure in such YBCO films
find that the correlated pinning centers in question
are in fact lines of dislocations that 
appear naturally in the PLD growth process\cite{klaassen}\cite{huij}.
The pinning centers notably arrange themselves in a manner that resembles
a snapshot of a two-dimensional (2D) liquid, 
as opposed to a snapshot of a 2D gas.

Below, we 
explore the theoretical consequences of the proposal
that the vortex lattice induced
by  an external magnetic field oriented near the $c$ axis  in 
films of 
PLD-YBCO
is in a thermodynamic Bose glass state
characterized by an infinite tilt modulus\cite{N-V}.
We specifically focus on the high-field regime
at the extreme type-II limit,
in which case  only a fraction of the vortex lines 
are localized at the dislocations 
that thread the film along the $c$ axis\cite{IL}\cite{nono}\cite{jpr04c},
and in which case the pinning of the
vortex lattice is collective\cite{LO}\cite{kes}\cite{M-E}.
The theory 
predicts a critical current density along the
film that approximately follows a
$1/\sqrt B$ power law as a function of magnetic field $B$
in the collective pinning regime.
It successfully accounts for the critical current density
obtained in $c$-axis field
from 
films of PLD-YBCO that are microns thick,
at liquid nitrogen temperature.
It fails, however, for much thinner films
at lower temperature. 
We demonstrate that this failure
can be corrected by including the effect of
point pins on the interstitial vortex lines that lie in between the 
correlated pins.  
They contribute an inverse dependence  on   film thickness
to the 
the critical current density
in magnetic field oriented near the  $c$ axis.

Consider  again the  
film of YBCO just described
in large enough external magnetic field  applied along the $c$ axis
so that magnetic fluxlines overlap considerably.
%
The vortex lines  experience  long range repulsive interactions
in such case, 
which favors their arrangement into a triangular lattice.
The natural material line defects that also run parallel to the $c$ axis
act to pin down
some fraction of the vortex lines,
on the other hand.
This frustrates the formation of a vortex lattice.
The resulting thermodynamic state is a Bose glass\cite{N-V}
that displays a divergent tilt modulus
because of pinning by the 
material line defects\cite{IL}\cite{nono}\cite{jpr04c}.

The vortex lattice described above is effectively 2D due to  the
strong orientation
along the $c$ axis.
Theoretical and numerical  calculations then indicate
that a net concentration of straight
lines of  unbound dislocations  will be quenched
into the triangular vortex lattice at zero 
temperature\cite{M-E}\cite{zeng}\cite{N-S}.
Monte Carlo simulations of 
the frustrated $XY$ model
further show that the dislocations 
in the vortex lattice appear either
unbound or  bound-up into neutral pairs
in the extreme type-II limit\cite{nono}.
In particular, the lines of dislocations do {\it not} arrange themselves
into low-angle grain boundaries
(cf. refs \cite{chandran} and \cite{menghini}). 

We shall now compute the critical current density that is  expected
along the film direction
for the hexatic Bose glass described above\cite{nono}\cite{jpr04c}.
The vortex lattice 
is pinned collectively\cite{LO} 
when the number of vortex lines localized at
material line defects is small
compared   to the total number of vortex lines.
This shall be assumed throughout.
The positional coherence of the vortex lattice along the $c$ axis implies
2D collective pinning in particular\cite{LO}\cite{kes},
with an infinite Larkin length in that
direction\cite{jpr04c}: $L_c \rightarrow \infty$.
Next, we shall assume that the critical current density
is limited by {\it plastic creep}
of the vortex lattice due to slip of the quenched-in
lines of  dislocations
along their respective glide planes\cite{book}.
The  Larkin length
in the direction transverse to the magnetic field, $R_c$, 
is then obtained by minimizing the sum of the elastic energy cost
and the gain
in pinning energy due to the translation of a Larkin domain
by an elementary Burgers vector 
of the triangular vortex lattice,
$b = a_{\triangle}$. 
This yields the estimate\cite{LO}\cite{kes}
\begin{equation}
R_c^{-2}   =
  C_0^2 n_{\rm p} (f_{\rm p} / c_{66} b)^2,
\label{ratio}
\end{equation}
for the density of Larkin domains,
where $n_{\rm p}$ denotes
the density of pinned vortex lines,
where $f_{\rm p}$ denotes the
maximum pinning force per unit length along a  material line defect,
and where
$c_{66}$ denotes the elastic shear modulus.
The prefactor $C_0$ above  is of order unity.
On the other hand,
because the Larkin correlation volume corresponds to the largest bundle
of vortex lines   that exhibits a purely elastic  response\cite{LO}, 
it is natural to identify
the average separation between lines of dislocations 
that are quenched into the vortex lattice
with the transverse Larkin length, $R_c$.
A similar minimization
of the sum of the elastic energy and
the  pinning energy
then yields the same form (\ref{ratio}) for the density of Larkin domains
in such case\cite{M-E},
but    with a prefactor
$C_0 \cong  \pi /   {\rm ln} (R_c / a_{\rm df}^{\prime})^2$.
Here  $a_{\rm df}^{\prime}$ is of order the
core diameter of a dislocation in the vortex lattice.
Finally,
the statistical nature of 2D  collective pinning
requires many pinned vortex lines 
per Larkin domain\cite{LO}\cite{Tink}: $n_{\rm p} > R_c^{-2}$.
Comparison with relation (\ref{ratio}) 
then yields  the  threshold in magnetic field 
%
\begin{equation}
B_{\rm cp} = C_0^2
(\sqrt{3} / 2)
(4 f _{\rm p} / \varepsilon_{0})^2 
\Phi_0,
\label{b_cp1}
\end{equation}
above which 2D collective pinning holds.  
The estimate
$c_{66} = (\Phi_0 / 8 \pi \lambda_L)^2 n_{\rm vx}$
for the shear modulus of the vortex lattice\cite{brandt}
has been  used here,
where  $n_{\rm vx}$  denotes the density of vortex lines,
where $\lambda_L$ denotes the London penetration depth,
and where $\Phi_0$ denotes the flux quantum.
Above, 
$\varepsilon_0 = (\Phi_0/4\pi \lambda_L)^2$ is the maximum tension
of  a fluxline in the superconductor.

The Lorentz force balances the pinning force
in the critical state 
following
$j_c  B / c  =  ( R_c^{-2}   \cdot  n_{\rm p} )^{1/2} f_{\rm p}$
when 2D  collective pinning holds\cite{LO}.
Here  $j_c$ denotes the critical current density along the film,
which is perpendicular to the magnetic field $B$
aligned parallel to the material line defects. 
Substitution of the result (\ref{ratio}) for the density of Larkin domains
reduces this balance
to
\begin{equation}
j_c  B / c  =  
C_0
(c_{66} b)^{-1} n_{\rm p} f^2_{\rm p}.
\label{balance2}
\end{equation}
The critical current density can then be determined from
the above condition for
mechanical equilibrium 
once the density of pinned vortex lines, $n_{\rm p}$,  is known.
Non-interacting vortex lines, $c_{66}/f_{\rm p}\rightarrow 0$,
yields the upper bound for $n_{\rm p}$.
The profile of  $n_{\rm p}$ versus
magnetic field 
in this case is clearly  just  the upper pair of dashed lines
that join at the density of material line defects, $n_{\phi}$,
shown in Fig. \ref{n_p}.
Infinitely weak correlated pins, $c_{66}/f_{\rm p}\rightarrow\infty$,
yields  the lower bound for $n_{\rm p}$, on the other hand.
It corresponds to the limit of an infinitely rigid vortex lattice.
The profile of  $n_{\rm p}$ versus
magnetic  field in that case follows 
the lower dashed line shown in Fig. \ref{n_p},
where $r_{\rm p}$ denotes the effective pinning radius of the material
line defects.
It is obtained by observing that
$n_{\rm p} / n_{\phi}$ 
coincides with
the probability that a
given
pinning center trap a vortex line,
$p = \pi r_{\rm p}^2 \cdot  n_{\rm vx}$.
(See Fig. \ref{vl} and ref. \cite{single}.) 
This coincidence is true for {\it any} density of pinned vortex lines,
$n_{\rm p}$, as long as the material line defects do not
crowd together: $\pi r_{\rm p}^2\cdot n_{\phi}\ll 1$.
Such is 
the case 
with 
the lines of dislocations that thread  films of PLD-YBCO,
which display an effective
hard-core repulsion\cite{klaassen}\cite{huij}.

The fact that the vortex lattice  is 
well ordered within
a Larkin domain 
(see Fig. \ref{vl})
indicates that the direct scaling of $n_{\rm p}$ with 
$B$ obtained  above in the limit of an infinitely rigid
vortex lattice
 persists 
in the regime of  weak correlated
pinning
more generally.
This in fact can be demonstrated by first observing
that the relative statistical error in the 
number of pinned vortex lines
at $f_{\rm p}\rightarrow 0$
obeys the
law of large numbers:
$\Delta N_{\rm p}/ {\bar{N_{\rm p}}} = 
f(\pi r_{\rm p}^2\cdot n_{\rm vx})/\sqrt{n_{\phi} R_c^2}$,
where $N_{\rm p}$ denotes the number of pinned vortex lines
inside of a Larkin domain,
and where $f(p) = \sqrt{(1-p)/p}$ 
sets the fluctuation scale 
specifically for the
binomial probability distribution\cite{reif}.
Second,
adapting the statistical principle of Larkin-Ovchinnikov\cite{LO}
to the present case
yields the new rule
that the break-up of the pristine
vortex lattice into Larkin domains of dimensions $R_c \times R_c$	
serves to increase the number of pinned vortex lines
with respect to the most probable number,
$\bar{N_{\rm p}} = p\cdot n_{\phi} R_c^2$
at $f_{\rm p}\rightarrow 0$,
by a number of order the statistical error, $\Delta N_{\rm p}$:
%
\begin{equation}
{\rm ln}
\Biggl({n_{\rm p} / n_{\phi}\over{\pi r_{\rm p}^2 \cdot n_{\rm vx}}}\Biggr)
 =   c_0 {\Delta N_{\rm p}\over{\bar N_{\rm p}}}.
\label{prob}
\end{equation}
%
Here $c_0$ is a numerical constant of order unity.
The density of pinned vortex lines 
therefore scales nearly directly with magnetic field
at small relative errors, $\Delta N_{\rm p} / \bar N_{\rm p} \ll   1$.
This requires collective pinning, $\bar N_{\rm p}\gg 1$.
Substitution 
into Eq. (\ref{ratio})
yields a density 
of Larkin domains,
$R_c^{-2} 
\cong (\sqrt{3}\pi/2) C_0^2
  (4f_{\rm p} r_{\rm p}/\varepsilon_0)^2 n_{\phi}$,
that depends  only weakly on  the magnetic field
in such case.
Finally,
the  function $f(p)$ specific to the binomial probability
distribution
can be approximated by 
$(2/\sqrt{3})\,{\rm ln} [(3\sqrt{e}/4)/p]$
for probabilities $p$ inside of the range
$[0.01, 0.98]$ (cf. ref. \cite{single}).
Substituting that approximation into Eq. (\ref{prob})
then yields the scaling result
\begin{equation}
n_{\rm p}  / n_{\phi} =  (B / B_{\phi 2})^{d_{\rm p} / 2}
\label{ansatz}
\end{equation}
for the density of pinned vortex lines
as a function of magnetic field
oriented parallel to the material line defects,  
where
$d_{\rm p} / 2 =  1 -  (c_1/\sqrt{n_{\phi} R_c^2})$ 
gives one-half the effective scaling dimension,
and  where
$B_{\phi 2} = c_2 \Phi_0/\pi r_{\rm p}^2$ 
gives the saturation field.
Here, we define
$c_1 = 2 c_0 / \sqrt{3}$
and $c_2 = (3\sqrt{e}/4)^{- [(2/d_{\rm p}) - 1]}$.
The last constant is notably less than unity. 
Fig.    \ref{n_p} depicts the field dependence of the
scaling ansatz (\ref{ansatz}) schematically,
where the correction to the scaling dimension is ignored. 

Table \ref{limits} lists various limiting cases that obey 
the scaling ansatz (\ref{ansatz}) for
the density of pinned vortex lines
as a function of magnetic field,
including the lower limit at weak pinning just treated.
Substituting (\ref{ansatz})
into the condition (\ref{balance2}) for the critical state
then yields the power law 
\begin{equation}
j_c (B) / j_c (B_{0}) =
({B_{0} / B})^{\alpha_{\rm cp}}
\label{powerlaw}
\end{equation}
for the critical current density as a function of magnetic field,
with exponent
\begin{equation}
\alpha_{\rm cp} = (3 - d_{\rm p}) / 2.
\label{alpha1}
\end{equation}
The reference critical current density
$j_c (B_0)$  
at the geometric mean 
$B_0 = (B_{\rm cp} B_{\phi 2})^{1/2}$
of the range in magnetic field of the power law,
$[B_{\rm cp}, B_{\phi 2}]$,
determines the latter
by the formula
\begin{equation}
{B_{\phi 2}\over{B_{\rm cp}}} = 
\Biggl(
C_1^{-1}
\cdot
{B_{\phi} \over{\sqrt{B_0 H_{c2}}}}
\cdot
{j_0\over{j_c (B_0)}}
\Biggr)^{\bigl[1 -  {1\over 2}(\alpha_{\rm cp} - {1\over 2})\bigr]^{-1}}.
\label{range}
\end{equation}
Here
$j_0 = 4 c \varepsilon_0  / 3\sqrt 3  \Phi_0\xi$ and 
$H_{c2} = \Phi_0 / 2\pi\xi^2$ are the 
the depairing current density
and the upper critical field, respectively,
each  set by the coherence length $\xi$ \cite{Tink}. 
Also, the constant  factor above is defined by
$C_1 =  (16\sqrt{\pi} / 3^{5/4}) C_0$.
Study of Eqs. (\ref{ratio}) and (\ref{prob}) 
again shows that $n_{\rm p}$ scales nearly directly with $B$
in the 	collective pinning regime,
$B  >   B_{\rm cp}$.
By the previous
 Eqs. (\ref{ansatz})-(\ref{alpha1}),
we conclude 
that the critical current density then decays with magnetic field like
$1/\sqrt B$ 
(see Table \ref{limits}).

Consider next the addition of a field  of material point defects
to the thin-film superconductor.
Either the Bose glass state described above 
remains intact\cite{N-V},  
in which case $L_c$ remains infinite\cite{jpr04c},
or it will break up into smaller Larkin domains of
finite thickness, $L_c < \infty$.
In either case, consider henceforth films of thickness
$\tau < L_c$.
Collective pinning
therefore remains 2D\cite{LO}\cite{kes},
and the {\it interstitial} vortex lines 
that remain free of the material line defects
may be considered as rigid rods.
The  material  point defects
will pin the latter.
Recall now 
the basic idea behind  2D collective pinning,
which  is that the critical pinning 
force per unit volume 
is given by the quotient of the
variance of the  net equilibrium force per unit length
over a Larkin domain
with 
 the
 cross-sectional area
 $R_c^2$.
The magnitude-square of this quotient   is equal to
\begin{equation}
\overline{|\sum_{i\in \rm{LD}}\vec f (i)|^2}/R_c^4
= (n_{\rm p} f_{\rm p}^2 + n_{\rm p}^{\prime} f_{\rm p}^{\prime 2})/R_c^2.
\label{mean-square}
\end{equation}
Above, $\vec f$ denotes the force per unit length
due either to a vortex line  pinned by a material line defect
or  to an interstitial vortex line pinned by material point defects.
These      have respective root-mean-square values of
$f_{\rm p}$ and $f_{\rm p}^{\prime}$.
Also,
$n_{\rm p}$ and $n_{\rm p}^{\prime}$ 
denote,
respectively, 
the density   of vortex lines
pinned by material line defects 
and the density of interstitial vortex lines pinned by material point defects. 
Last, the overbar notation above denotes a bulk average 
achieved by rigid translations
of a given Larkin domain (LD).
The identity (\ref{mean-square}) is then  due to 
the statistical independence of all of 
the pinning forces: 
$\overline{\sum_{i,j\in{\rm LD}}^{\prime}\vec f (i) \cdot \vec f (j)} = 0$, 
where the prime over the summation symbol specifies that
$i$ and $j$ denote distinct pinned vortex lines
within a given Larkin domain.
This last property is closely related to the null  average
value of the net pinning
force exerted on a Larkin  domain
when the  vortex lattice is in mechanical equilibrium.
In particular,
$\sum \vec f  = 0$
implies
$\overline{\sum_{\rm{LD}}\vec f} = 0$.
By Eq.  (\ref{mean-square}),
the   additional source of pinning due to the interstitial
vortex lines  can then be accounted 
for by making the replacements
\begin{equation}
n_{\rm p} f_{\rm p}^2\rightarrow
n_{\rm p} f_{\rm p}^2 + n_{\rm p}^{\prime} f_{\rm p}^{\prime 2}
\quad {\rm and} \quad
j_c \rightarrow j_c + j_c^{\prime}
\label{replace}
\end{equation}
in Eqs.   (\ref{ratio}) and (\ref{balance2}) 
for the density of Larkin domains in the vortex lattice 
and for the critical state. 
Here, $j_c^{\prime}$ denotes the contribution by the point pins to the
net critical current density, $J_c = j_c + j_c^{\prime}$.
The first replacement 
displayed by 
Eq. (\ref{replace})
indicates an {\it effective} density of vortex lines pinned by
material line defects, 
$n_{\rm p} + (f_{\rm p}^{\prime 2} / f_{\rm p}^2) n_{\rm p}^{\prime}$.
Notice that the latter correctly equals $n_{\rm p}$
when $f_{\rm p}^{\prime} = 0$,
and $n_{\rm p} + n_{\rm p}^{\prime}$
when $f_{\rm p}^{\prime} =f_{\rm p}$.
Matching it to the modified result for the density of Larkin domains
(\ref{ratio}) then yields the old result (\ref{b_cp1}) for
the  threshold field  beyond which 2D collective-pinning holds.
The previous
Eqs. (\ref{balance2}) and (\ref{range})
 then remain unchanged 
as long as $j_c$ is understood
to represent the contribution to the critical current density
by the material line defects alone!  
 
The condition ({\ref{balance2}) for the critical state
that results from pinning of the vortex lattice by  material line defects
demonstrates that
the  critical current density 
 $j_c$ is a {\it bulk} quantity
that is independent of the thickness of the thin-film superconductor
along the axis of the  correlated pins. 
This notably is not the case for the  contribution by point pins 
to the critical current density, $j_c^{\prime}$.
In particular, the forces due to point pins 
add up statistically along a   rigid interstitial vortex line
if the film is much thicker than 
the average separation between such pins along the field direction,
$\tau_{\rm p}^{\prime}$.
The effective pinning force per unit length experienced by an interstitial
vortex line  is then given by\cite{kes}
$f_{\rm p}^{\prime} =  
f_0^{\prime} /(\tau_{\rm p}^{\prime} \tau )^{1/2}$
at film thicknesses $\tau > \tau_{\rm p}^{\prime}$,
where $f_0^{\prime}$ denotes the maximum force exerted by a point pin.
Also, all of the interstitial vortex lines are pinned by point defects
if the film is thick enough:
$n_{\rm p}^{\prime} = n_{\rm vx} - n_{\rm p}$ at  
$\tau > \tau_{\rm p}^{\prime}$. 
The relative 
contribution by point pins to the critical current density
is then predicted to show an inverse dependence 
on
film thickness,
$j_c^{\prime} / j_c = \tau_0 /\tau$,
that is set by the scale
$\tau_0 = 
(n_{\rm vx}/n_{\rm p} - 1)(f_0^{\prime 2}/f_{\rm p}^2 \tau_{\rm p}^{\prime})$.
The  net critical current density, $J_c = j_c + j_c^{\prime}$,
then follows a
pure $1/\sqrt B$ power law 
in the case that
the density of vortex lines pinned by 
material line defects scales directly
with magnetic field
($d_{\rm p} = 2$).
More generally,
we predict a  linear dependence on film thickness for the
net critical current per unit width
following
$K_c = (\tau_0 + \tau) j_c$, 
 where
$K_c = \tau J_c$ by definition.


The summation of pinning forces is
{\it coherent}, on the other hand,
at magnetic 
fields below the threshold (\ref{b_cp1})  for 2D collective pinning:
$j_c  B / c  =   n_{\rm p} f_{\rm p}$
at $B < B_{\rm cp}$.
Further, we have that
$n_{\rm p}$ is approximately equal to 
$n_{\rm vx}$
at low  magnetic fields compared to the accommodation
scale\cite{N-V} 
$B_{\phi 1} = (4 \varepsilon_{\rm p} / \varepsilon_0) 
(\Phi_0\cdot n_{\phi})$,
where $\varepsilon_{\rm p}$ denotes the depth of the correlated
pinning potential per unit length.
The above balance of forces 
then yields
a critical current density that
reaches a plateau 
%
\begin{equation}
j_c (0+) / j_0 =
(3 \sqrt{3} /4)
(f_{\rm p} \xi/\varepsilon_0)
\label{j_c3}
\end{equation}
in the zero-field limit\cite{civale91}.
Finally, 
the inequality $B_{\phi 1} < B_{\rm cp}$ shall be assumed throughout.
It places the bound
$B_{\phi} < (4 \varepsilon_{\rm p} / \varepsilon_0) (\Phi_0/r_{\rm p}^2)$
on the matching field.

Thin films of YBCO  grown by PLD on a
substrate 
also show an extended regime in magnetic field
oriented   parallel  to the $c$ axis
where the critical current
obeys a  power law (see Fig. \ref{jcvsb} 
and refs. \cite{dam}\cite{klaassen}\cite{suenaga}\cite{maiorov}).
The films themselves
are divided into columns of subgrains that run
parallel to the crystallographic $c$ axis.
Etching of the 
film surface 
demonstrates that lines of dislocations
also run parallel to
the $c$ axis
in ``trenches'' that separate such growth islands\cite{klaassen}\cite{huij}.
There typically exists about one threading dislocation
per growth island, each separated
by a distance of about $110$ nm \cite{klaassen}\cite{huij}.
The dimensionality of the columnar pins
along a cross section
then coincides with 
that of the subgrains, $d_{\rm p} = 2$.
The critical current 
that is obtained
from such films of YBCO
in magnetic field $B$ aligned along the $c$-axis 
typically obeys 
a $1/\sqrt B$ law, 
in particular\cite{dam}\cite{klaassen}\cite{suenaga}\cite{maiorov}.
This is consistent with a direct dependence,
$n_{\rm p}\propto B$,
of the density of pinned vortex lines 
on the magnetic field
by the present theory (\ref{alpha1}}) 
for 2D collective pinning.
Such a dependence on   magnetic field,
in turn,
is predicted approximately by Eq. (\ref{prob})
in the collective pinning regime.
(See
Table \ref{limits}).

Another point of comparison between theory and experiment is the
actual magnitude of the critical current density.
The maximum force per unit length
exerted by a correlated pin, $f_{\rm p}$, can be obtained from $j_c$
in self field via Eq. (\ref{j_c3}) \cite{c-m}.
Substituting it
into the right-hand side of Eq. (\ref{b_cp1})  
then yields a prediction for the  threshold field
beyond which 2D collective pinning holds:
%
\begin{equation}
B_{\rm cp} = 
C_1^{2}
[j_c (0+) / j_0]^2
H_{c2}.
\label{b_cp2}
\end{equation}
Values of self-field $J_c$
measured in various 
films of PLD-YBCO\cite{klaassen}\cite{suenaga}\cite{maiorov}
are listed in Table \ref{fits},
along side  of the  predicted threshold field, $B_{\rm cp}$. 
The London penetration depth is set to its value at zero temperature,
$\lambda_L (0) = 150$ nm.
A value for the  superconducting coherence length of $\xi (0) = 1.5$ nm
then yields a depairing current density of $j_0 (0) = 300$ MA/cm$^2$,
as well as an
upper-critical field of $H_{c2}  (0) = 146$ T.
Also,
the prefactor $C_0$ 
on the right-hand side of Eq. (\ref{ratio}) 
for the density of Larkin domains in the vortex lattice is set to unity.
The predicted threshold magnetic field
fails to  lie below the geometric mean $\sqrt{B_1 B_2}$ of
the power-law regime
(see Fig. \ref{jcvsb}) only in the case of 
the relatively thin YBCO film\cite{klaassen}\cite{suenaga}\cite{maiorov}.
This is likely due to  the neglect of pinning by 
true grain boundaries\cite{diaz}.

The present theory for 2D collective pinning by material line defects
also predicts the range of the power law in magnetic field (\ref{range}) 
from knowledge of the critical current density 
at the logarithmic midpoint. 
 (See Fig. \ref{jcvsb}, ``$\times$''.)
Table \ref{fits} again lists
recent measurements of the critical current density
in films of PLD-YBCO
at the logarithmic midpoint,
along side of the predicted range.
Only films that show an inverse power-law
(\ref{powerlaw}) in $J_c$ versus $B$
characterized by an exponent $\alpha$
 in the vicinity of $1/2$ 
are considered.
Standard physical parameters for YBCO are again used,
the prefactor $C_0$ is again set to unity,
and $\varepsilon_0$ is  again set to its value at zero temperature.
Additionally, 
the effect of point pins is neglected ($j_c^{\prime} = 0$).
Comparison of the last two columns in Table {\ref{fits}
indicates that the present theory for 2D collective pinning
gives a fair
account of the critical current density in 
PLD films of YBCO
superconductor that are microns thick.
We believe that the failure of the theory in the case of the much thinner film
is due to the neglect of the contribution by point pins.



In conclusion, 
2D collective  pinning of the vortex lattice by material line 
defects can account for the inverse-square-root
power law obeyed by the critical
current density in films of PLD-YBCO 
versus external magnetic field\cite{dam}.
We also predict on this basis that the critical current per
unit width is a  linear  function of film thickness,
with a positive slope equal to the bulk critical current density,
and with a negative
intercept
on  the thickness axis,
after extrapolation,
due to pinning of interstitial
vortex lines by material point defects.

\acknowledgments  The authors thank L. Civale, S. Foltyn, H. Fertig,
 A. Koshelev, B. Maiorov, C. Olson, V. Vlasko-Vlasov and J. Willis
for valuable discussions.
JPR is indebted to E. Rezayi for finding an error
in an earlier version of the manuscript
concerning the ``law of large numbers''.
This work was performed 
under the auspices of the U.S. Department of Energy.

\begin{figure}
\includegraphics[scale=0.31, angle=-90]{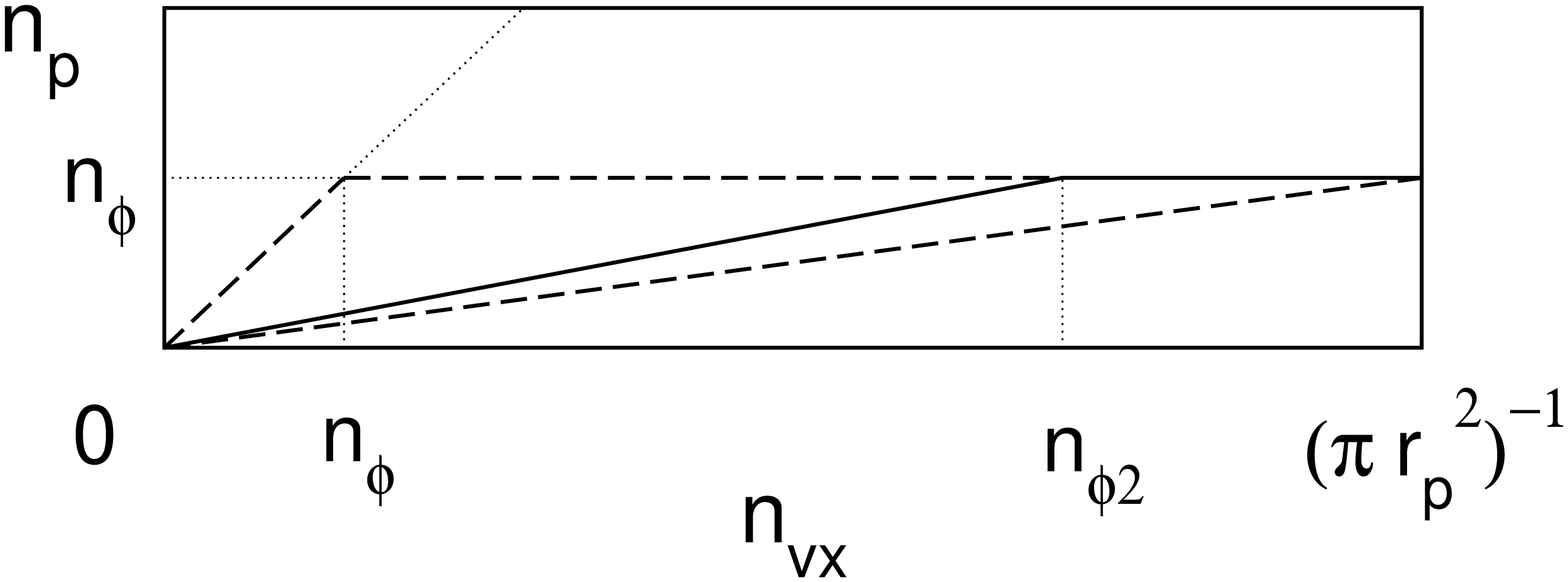}
\caption{Sketched is the density of vortex lines pinned to material line
defects versus the net density of vortex lines.
The scale is assumed to be large compared to
the accommodation scale ($B_{\phi 1}$), 
which is not shown.
(See text and ref. \cite{N-V}.)} 
\label{n_p}
\end{figure}

\begin{figure}
\includegraphics[scale=0.31, angle=-90]{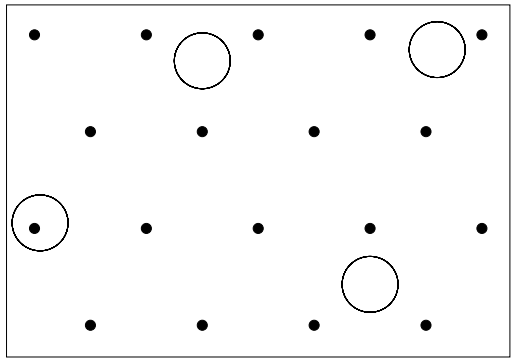}
\caption{Sketched is a portion of the vortex lattice (dots) 
found inside of a Larkin domain,
with a random arrangement of linear pinning centers (circles).}
\label{vl}
\end{figure}

\begin{figure}
\includegraphics[scale=0.31, angle=-90]{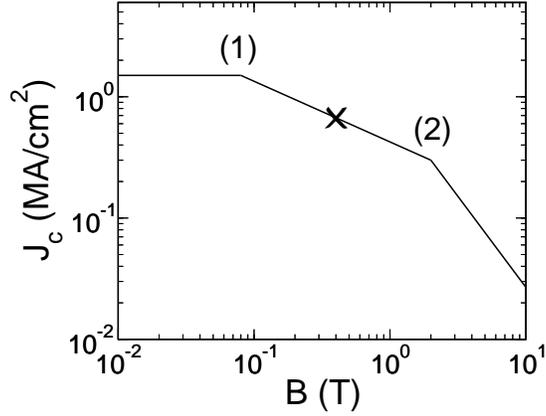}
\caption{Sketched is a typical profile
of the critical current density measured
in films of PLD-YBCO
as a function of external magnetic field oriented near the $c$ axis.
The ``$\times$''
 marks the power-law regime
 in magnetic field,
[$B_1, B_2]$.}
\label{jcvsb}
\end{figure}

\begin{table}
\begin{center}
\begin{tabular}{|c|c|c|c|}
\hline
 material line defects &  $d_{\rm p}$
& $B_{\phi 2}$ 
& $\alpha_{\rm cp}$ 
\\
\hline
saturated: $n_{\rm p} = n_{\phi}$  & $0$ & --
& $3/2$ 
 \\
\footnote{at commensuration.} row,  period $D$ & $1$
& $\Phi_0 / (\sqrt 3 / 2) D^2$ 
&  $1$
 \\ 
$^a$ triangular lattice &  $2$  & $\Phi_0 \cdot n_{\phi}$  
& 1/2
  \\
\footnote{with hard-core repulsion 
(see text and refs. \cite{klaassen} and \cite{huij}).}
random, $c_{66}/f_{\rm p}\rightarrow\infty$ & $2$
& $\Phi_0/\pi r_{\rm p}^2$ 
& $1/2$
\\
\hline
\end{tabular}
\caption{Listed are various examples of  pinning by
material line defects that obey
the scaling ansatz (\ref{ansatz}) for the density of pinned vortex lines as a 
function of magnetic field.}
\label{limits}
\end{center}
\end{table}

\begin{table}
\begin{center}
\begin{tabular}{|c|c|c|c|c|c|c|c|c|c|c|}
\hline
 Laboratory
 & thick &  $T$  & $J_c (0+)$ & $B_{\rm cp}$
&  $\alpha$ & $B_{\phi}$ & $\sqrt{B_{1} B_{2}}$
& $J_c (\sqrt{B_{1} B_{2}})$ & $B_{\phi 2}/B_{\rm cp}$ & $B_{2}/B_{1}$ \\
\hline
Amsterdam, ref. 
\cite{klaassen} & $0.1 \mu$m &  $40$ K & $10$ MA/cm$^2$  & $8.4$ T
& $0.64$ & $0.07$ T 
& $0.78$ T & $2.3$ MA/cm$^2$ & $0.1$ & $6.7$ \\
LANL, ref. 
\cite{maiorov} & $4.3 \mu$m  & $75.5$ K & $1.0$ MA/cm$^2$  & $0.08$ T
& $0.46$ & 
\footnote{Rough estimate taken from ref. \cite{huij}.}
$0.17$ T
 & $0.31$ T & $0.4$ MA/cm$^2$ & $2.9$  & $20$ \\ 
BNL, ref. 
\cite{suenaga} & $3 \mu$m &   $77$ K & $1.3$ MA/cm$^2$  & $0.14$ T
& $0.67$  & $^{\it a}$ $0.17$ T & $0.22$ T & $0.3$ MA/cm$^2$ & $4.7$ & $20$ \\
\hline
\end{tabular}
\caption{Listed above are theoretical predictions,
[$B_{\rm cp}, B_{\phi 2}$],
versus experiment, $[B_1, B_2$]. (See Fig. \ref{jcvsb}.)
The effect of point pinning is neglected
($j_c^{\prime} = 0$).}
\label{fits}
\end{center}
\end{table}


\begin{thebibliography}{}
\bibitem{single} A material line defect
with an effective circular crossection
will begin to trap  two vortex lines in the 
pristine vortex lattice
at probabilities $p = \pi r_{\rm p}^2 / (\sqrt{3}/2) a_{\triangle}^2$ 
larger than
$\pi/2\sqrt{3}\cong 0.907$,
where $r_{\rm p} = a_{\triangle} / 2$.
\end{thebibliography}
\end{document}